\documentclass[prb,aps,floats,amssymb,showkeys,showpacs,preprint,
superscriptaddress,tightenlines]{revtex4}
\usepackage{graphicx}
\usepackage{epsfig,bm}
\begin{document}
\preprint{Bicocca-FT-xx-yy  June 2007}

\title{ High-temperature expansions through order 24 
for the two-dimensional classical XY model on the square lattice}
\author{P. Butera\cite{pb}}
\affiliation
{Istituto Nazionale di Fisica Nucleare and\\
Dipartimento di Fisica, Universit\`a di Milano-Bicocca\\
 3 Piazza della Scienza, 20126 Milano, Italy}

\author{M. Pernici\cite{mp}}
\affiliation
{Istituto Nazionale di Fisica Nucleare and\\
Dipartimento di Fisica, Universit\`a di Milano\\
 16 Via Celoria, 20133 Milano, Italy}

\date{\today}
\begin{abstract}
The high-temperature expansion of the spin-spin correlation
 function of the two-dimensional classical XY (planar rotator) 
 model on the square lattice is extended 
 by three terms, from order 21 through order 24, and analyzed to improve the 
estimates of the critical parameters.
\end{abstract}

\maketitle

Tests of increasing accuracy~\cite{Kenna} of the 
BKT theory~\cite{BKT} 
of the two-dimensional XY model critical behavior have been made possible
 by the steady improvements of the computers performances and  
the progress in the  numerical approximation algorithms.  
However, the critical parameters of this model 
have not yet been determined with a precision comparable 
to that reached for the usual power-law critical phenomena,
 due to the complicated and peculiar nature of the 
critical singularities.
 Therefore any effort at improving the accuracy of the 
available numerical methods by   stretching them  towards their (present) 
limits should be welcome.
 After extending the high-temperature(HT) expansions
 of the model in successive steps~\cite{BC}  from order 
$\beta^{ 10}$ to $\beta^{ 21}$, we present here a further 
 extension by three orders for the expansions of the spin-spin correlation 
on the square lattice  and perform a first brief analysis of
 our data for the susceptibility and the second-moment
 correlation-length. More results and further extensions
 both for the square and  the triangular lattice\cite{bct}
 will be presented elsewhere. Our study strengthens the  support
of the main results of the BKT theory  already coming from the analysis 
 of shorter series   and  suggests a closer agreement with  recent 
high-precision simulation studies\cite{Kenna,Has}  of the model.

The    Hamiltonian

\begin{equation}
H\{ v \} = - 2{J}  \sum_{nn} 
\vec v({\vec r}) \cdot \vec v({\vec r}')
\label{hamilt}\end{equation}
 with $\vec v({\vec r})$ 
 a two-component unit vector  at the  site ${\vec r}$ of a square lattice,
describes a system of $XY$ spins  with nearest-neighbor interactions.

Computing  the spin-spin correlation function, 
\begin{equation}
C(\vec 0, \vec x;\beta)= <s(\vec 0) \cdot s(\vec x)>,
\label{corfun}
\end{equation}  
(for all values of $\vec x$ for which the HT expansion coefficients are 
non-trivial within the maximum order reached),  
as series expansion in the  variable  $\beta= J/kT$,
 enables us to evaluate  the  expansions of the  $l$-th order 
spherical moments  of the correlation function:
\begin{equation}
m^{(l)}(\beta) = \sum_{\vec x  }|\vec x|^l <s(\vec 0) \cdot s(\vec x)>
\label{sfermom}
\end{equation}

and in particular the reduced ferromagnetic 
susceptibility $\chi(\beta)=m^{(0)}(\beta)$.
In terms of $m^{(2)}(\beta)$ and $\chi(\beta )$ we can form the 
 second-moment correlation length: 
\begin{equation}
\xi^2(\beta) = m^{(2)}(\beta)/4\chi(\beta).
\label{corleng}
\end{equation}

 Our results for the nearest-neighbor correlation function 
(or energy $E$ per link) are:

\begin{eqnarray}
E &=&
\beta  + \frac{3}{2}\beta^{3} + \frac{1}{3}\beta^{5} - 
 \frac{31}{48}\beta^{7} - \frac{731}{120}\beta^{9} - 
 \frac{29239}{1440}\beta^{11} - \frac{265427}{5040}\beta^{13} - 
 \frac{75180487}{645120}\beta^{15} 
\nonumber \\ &-& 
 \frac{6506950039}{26127360}\beta^{17}-  
 \frac{1102473407093}{2612736000}\beta^{19} - 
 \frac{6986191770643}{14370048000}\beta^{21}
\nonumber \\ &+&  
\frac{1657033646428733}{4138573824000}\beta^{23}+
   O(\beta^{25}) 
\nonumber \\
\end{eqnarray}

For the susceptibility we have:
\begin{eqnarray}
\chi &=&
  1 + 4\beta + 12\beta^{2} + 34\beta^{3} + 88\beta^{4} + 
 \frac{658}{3}\beta^{5} + 529\beta^{6} + \frac{14933}{12}\beta^{7} + 
 \frac{5737}{2}\beta^{8} + \frac{389393}{60}\beta^{9}  
\nonumber \\ &+& 
 \frac{2608499}{180}\beta^{10} + \frac{3834323}{120}\beta^{11} + 
 \frac{1254799}{18}\beta^{12} + \frac{84375807}{560}\beta^{13} + 
 \frac{6511729891}{20160}\beta^{14}  
\nonumber \\ &+& 
 \frac{66498259799}{96768}\beta^{15}+ 
 \frac{1054178743699}{725760}\beta^{16} + 
 \frac{39863505993331}{13063680}\beta^{17} 
 \nonumber \\ &+&
 \frac{19830277603399}{3110400}\beta^{18}+   
\frac{8656980509809027}{653184000}\beta^{19}+
 \frac{2985467351081077}{108864000}\beta^{20}
\nonumber \\ &+& 
 \frac{811927408684296587}{14370048000}\beta^{21}+ 
\frac {399888050180302157} {3448811520} \beta^{22}+
\frac {245277792666205990697} {1034643456000} \beta^{23}
\nonumber \\ &+&
\frac {83292382577873288741}{172440576000}\beta^{24}+
 O(\beta^{25}) 
\nonumber \\ 
\end{eqnarray}

For the second moment of the correlation function we have:
\begin{eqnarray}
m_2 &=& 
4\beta + 32\beta^{2} + 162\beta^{3} + 672\beta^{4} + 
\frac{7378}{3}\beta^{5} +\frac{24772}{3}\beta^{6} + 
\frac{312149}{12}\beta^{7} + 77996\beta^{8}
\nonumber \\ &+& 
\frac{13484753} {60}\beta^{9}+
\frac{28201211}{45} \beta^{10}+  
\frac{611969977}{360} \beta^{11}+ 
\frac{202640986}{45} \beta^{12}+   
\frac{58900571047}{5040}\beta^{13}
\nonumber \\ &+&
\frac{3336209179}{112}\beta^{14}+
\frac{1721567587879}{23040 }\beta^{15}+
\frac{16763079262169}{90720}\beta^{16}+
\frac{5893118865913171}{13063680}\beta^{17}
\nonumber \\ &+&
\frac{17775777329026559}{16329600}\beta^{18}+
\frac{1697692411053976387}{653184000}\beta^{19}+
\frac{41816028466101527}{6804000}\beta^{20}
\nonumber \\ &+&
\frac{206973837048951639371}{14370048000}\beta^{21}+
\frac{721617681295019782781}{21555072000}\beta^{22}
\nonumber \\ &+&
\frac{79897272060888843617033}{1034643456000}\beta^{23}+
\frac{2287397511857949924319}{12933043200}\beta^{24}+ O(\beta^{25})
\end{eqnarray}

 The coefficients of order less than 22 were  already tabulated in
Refs.\cite{BC}, but for completeness we  
report  all known terms.
As implied by eq.(\ref{hamilt}), 
the normalization of these series 
reduces to that of our earlier papers\cite{BC}
 by the change $\beta \rightarrow \beta/2$. 

Let us now list briefly the main predictions\cite{BKT} of the BKT
 renormalization-group analysis to which the HT series should
 be confronted in order to extract the critical parameters.

As $\beta \rightarrow \beta_{c}$, the correlation length 
$\xi^2(\beta) = m^{(2)}(\beta)/4\chi(\beta)$
 is expected to diverge  with the characteristic singularity
\begin{equation}
\xi^2(\beta) \propto \xi^2_{as}(\beta)= exp(b/\tau^\sigma)[1+O(\tau)]
\label{xias}
\end{equation}

where $\tau=1-\beta/\beta_{c}$.
The exponent $\sigma$ takes the universal value $\sigma=1/2$, whereas
$b$ is a nonuniversal positive constant.
At the critical inverse temperature $\beta=\beta_{c}$, 
the asymptotic behavior of the 
two-spin correlation function as $|\vec x |= r \rightarrow \infty$ 
is expected\cite{Amit} to be
\begin{equation} 
<s(\vec 0) \cdot s(\vec x)> \propto \frac {({\rm ln}r)^{2\theta}} {r^{\eta}}
[1+O(\frac{{\rm ln}{\rm ln}r} {{\rm ln}r})]
\label{coras}
\end{equation}
 Universal values $\eta= 1/4$ and $\theta = 1/16$ are predicted also 
for the correlation exponents.
 
A simple non-rigorous argument based on 
 eqs. (\ref{xias}) and (\ref{coras}) suggests that, for $l>\eta-2$,
 the spherical correlation moment $m^{(l)}(\beta)$ 
diverges as $\tau \rightarrow 0^+$ with the singularity 

\begin{equation}
m^{(l)}(\beta) \propto
\tau^{-\theta}\xi^{2-\eta+l}_{as}(\beta)[1+O(\tau^{1/2}{\rm ln}\tau)]
\label{momas}
\end{equation}

This argument was challenged\cite{Bal} by a recent renormalization group
 analysis implying that the logarithmic factor in  eq.(\ref{coras})
 gives rise to a less singular correction in the correlation moments,
 taking, for example in the case of the susceptibility, the form 

\begin{equation}
m^{(0)}(\beta) \propto
\xi^{2-\eta}_{as}(\beta)[1+cQ]
\label{momasb}
\end{equation}
     
where  $Q= \frac {\pi^2} {2({\rm ln}(\xi)+u)^2} +O({\rm ln}(\xi)^{-5})$ 
 and $u$ is a non universal parameter.

By eqs.(\ref{xias}) and (\ref{momas}), the ratios 
$r_n(m^{(l)})= a^{(l)}_n/a^{(l)}_{n+1}$ of the successive HT 
expansion coefficients of the correlation moment $m^{(l)}(\beta)$, 
 for large $n$ should behave\cite{BC} as 
\begin{equation}
r_n(m^{(l)})= \beta_{c} + C_l/(n+1)^{\zeta} +O(1/n)
\label{ratas}
\end{equation}

with $\zeta=1/(1+\sigma)$, to be contrasted with the value $\zeta=1$ 
 which is found for the  usual power-law critical singularities. 

\begin{figure}[tbp]
\begin{center}
\leavevmode
\includegraphics[width=3.37 in]{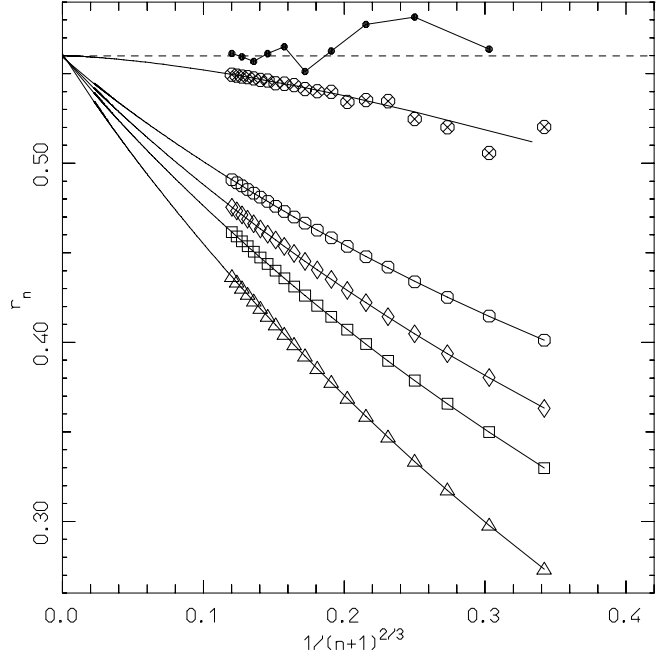}
\caption{ \label{figlograpp}  Ratios  of the 
successive HT-expansion coefficients vs. $1/(n+1)^{2/3}$: for 
the susceptibility $\chi$ (open circles),  
for $m^{(1/2)}$ (rhombs), for $m^{(1)} $ (squares) and
  for $m^{(2)}$ (triangles).
The four lowest continuous curves are  obtained  
by separate three-parameter fits of each ratio sequence to its
leading asymptotic behavior eq.(\ref{ratas}).
 The data points represented by crossed circles  are obtained by extrapolating 
 the sequence of the susceptibility alternate ratios  with respect to
 $1/n^{2/3}$, and the continuous line interpolating them 
is the result of a two-parameter fit of the last few points to
 the expected asymptotic form $a+b/n$. The small black circles are obtained by
 a further extrapolation of the latter quantities  with respect to $1/n$. 
 The continuous line interpolating the black circles is drawn only as a guide 
to the eye.
 The horizontal broken line indicates the critical value $\beta_c=0.55995$  
suggested by the simulation of Ref.\cite{Has}}
\end{center}
\end{figure}

To begin with, let us assume that $\sigma=1/2$ as expected, 
so that  $\zeta=2/3$.
Fig.\ref{figlograpp} gives a suggestive visual test of the 
asymptotic behavior of some ratio sequences $r_n(m^{(l)})$ 
by comparing them  with eq.(\ref{ratas}).
The four lowest   continuous curves interpolating the data points   
 are obtained by separate three-parameter fits of the ratio sequences 
  $r_n(\chi)$, $r_n(m^{(1/2)})$,  $r_n(m^{(1)})$ and $r_n(m^{(2)})$  to the 
asymptotic form $a +b/(n+1)^{2/3} +c/(n+1)$ of eq.(\ref{ratas}). 
In the same figure, 
 the two upper sets of points are obtained by  extrapolating the 
alternate-ratio sequence for the susceptibility, 
first in terms of $1/(n+1)^{2/3}$  and then in terms of $1/(n+1)$.
 The values of $\beta_c$ indicated by the fits of the ratio sequences,  
 range between 0.5592 and 0.5611. 

\begin{figure}[tbp]
\begin{center}
\leavevmode
\includegraphics[width=3.37 in]{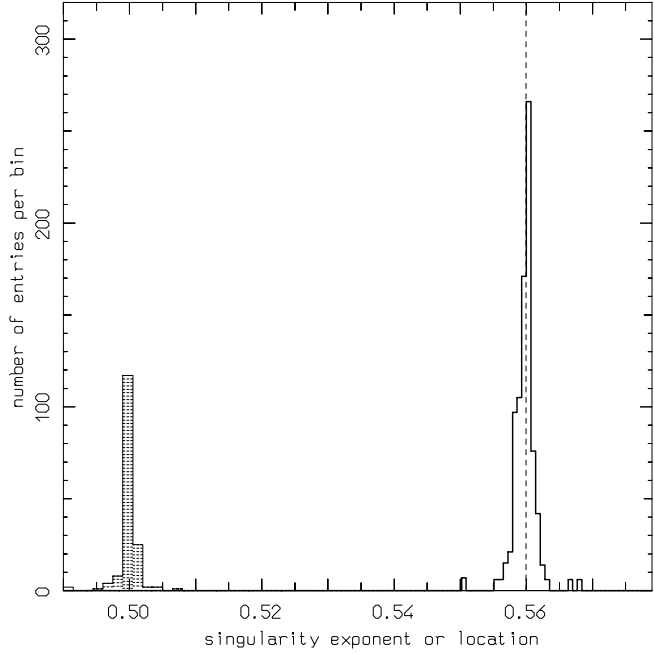}
\caption{\label{figbetacda2} Distribution  of singularities
 for a class of second-order inhomogeneous DAs 
 of $L(1.19,\beta)=(1.19+{\rm ln} \chi)^2$ versus their 
position on the $\beta$ axis(open histogram).
 The central value of the open histogram
 is $\beta_c=0.5598(10)$. The bin width is 0.0007.
  The vertical dashed line shows the critical value $\beta_c=0.55995$  
 indicated by the simulation of Ref.\cite{Has} for which one can guess 
an uncertainty at least twice smaller than ours. 
 The hatched histogram represents the distribution  of the exponent $\sigma$ 
 obtained from DAs of $L(1.19,\beta)$ biased with  $\beta_c= 0.5598$, vs.
 their position on the $\sigma$ axis.
The  central value of the hatched histogram  
is $\sigma=0.500(1)$ and the bin  width is 0.0015.
The variation of the central value of $\sigma$ as  $\beta_c$ varies in its
 uncertainty interval is $0.01$. This value can be taken as a more
reliable estimate of
 the uncertainty of  $\sigma$.}
\end{center}
\end{figure}

A more accurate analysis can be based on the simple remark that, 
near the critical point,
by eq.(\ref{xias}) and (\ref{momas}) (or eq.(\ref{momasb})),  
 one has ${\rm ln}(\chi) =c_1/\tau^{\sigma}+ c_2+..$.
Therefore, if $\sigma=1/2$, the relative strength of the $1/\sqrt{\tau}$
 and $1/\tau$ singularities in the function
 $L(a,\beta)=( a+ {\rm ln}(\chi))^{2}$ is determined by the value of the 
constant $a$. If we choose $a \approx 1.19$, the function  $L(a,\beta)$  
is approximately dominated by a simple pole and we can expect that the 
differential approximants (DAs)\cite{Gutt}
 will be able to determine with higher
 accuracy not only the position, but also the exponent of 
the critical singularity. 
 Using inhomogeneous second-order DAs of $L(a,\beta)$, we can 
locate the critical singularity at $\beta_c= 0.5598(10)$. 
 By analysing in the same way the  series data truncated to order 21 which
 were previously available , we would 
get the estimate $\beta_c=0.5588(15)$. A consistent estimate 
$\beta_c=0.558(2)$ had been obtained in  earlier independent\cite{BC,pisa}
 studies of the same series using  Pad\'e  approximants or first-order DAs.
Older studies\cite{BC} of slightly shorter series also indicated values of  
$\beta_c$ in the same range, but with notably larger uncertainty.
Thus our new series results indicate a stabilization 
 and a sizable reduction of the spread for the $\beta_c$ estimates.
 Our uncertainty estimates are generally taken as  the width of the 
distribution  of the  values of $\beta_c$ in the appropriate class of DAs. 
Fig.\ref{figbetacda2} shows the singularity distribution (open histogram) 
 of  the set of quasi-diagonal 
 DAs which yield our new estimate.
These are chosen as the approximants $[k,l,m;n]$   with $17 < k+l+m+n< 22$.
 Moreover, we have taken $|k-l|,|l-m|,|k-m|< 3$ with $ k,l,m > 3$ and $1<n<7$.
 The class of DAs can be varied with no significant variation of the
 final estimates, for example by further restricting the extent 
 of off-diagonality, or by varying the minimal degree of the polynomial 
coefficients in the DAs.  
No limitations have been imposed  on 
the exponents  of the singular terms or on the background terms in the 
DAs in order to avoid  biasing the $\beta_c$ estimates.
Should we require that the exponent of the most singular term in the
 approximants
 differs from -1,
 for example, by less than $20\%$, we would obtain $\beta_c=0.5602(5)$,
 well within the uncertainty of our previous unrestricted estimate.  
 The vertical dashed 
 line in  Fig.\ref{figbetacda2} shows the value  $\beta_c=0.55995$ 
 suggested by the simulation of Ref.\cite{Has}.
Although no explicit indication of an uncertainty comes with this estimate,  
 an upper bound to its error might be guessed from the statement\cite{Has} 
 that the simulation can exclude  values larger or equal than
 $\beta_c= 0.56045$ 
for the inverse critical temperature.

Biasing with  $\beta_c= 0.5598(10)$ the  set previously specified of
second-order DAs of $L(a,\beta)$, 
leads to the exponent estimate $\sigma=0.50(1)$.  
 Fig.\ref{figbetacda2} also shows  the distribution of the  exponent
estimates (hatched histogram) from  this biased set.  
 The uncertainty  we have reported for $\sigma$  accounts
 not only for the width of its distribution  
shown in   Fig.\ref{figbetacda2}, but   
also for the variation of its central value  as the  bias value of $\beta_c$ 
is varied in the uncertainty interval of the critical inverse temperature.
Essentially the same value of $\sigma$ would be obtained from the 
analysis of a series truncated to order 21.
 
While, as one should expect, the DA estimate of $\beta_c$ is rather insensitive
 to the choice of $a$, the estimate of the exponent $\sigma$ and the
 width of its distribution are fairly improved by our choice of $a$. Taking
 for example $a=0$, we would find $\sigma= 0.53(4)$, which shows how
 the convergence of the exponent estimates is slowed down 
by the more complicated singularity  structure of $L(0,\beta)$.
Similar values of $\sigma$ were found in previous studies of shorter
series. 
Probably for the same reason, also the central values of the  $\eta$ 
 estimates  obtained from the usual 
 indicators are still slightly larger than expected. 
For example, by studying the function
$ H(\beta)={\rm ln}(1+m^{(2)}/\chi^2)/\rm ln(\chi)$ (or analogous functions
 of different moments),  we can infer $\eta=0.260(10)$.
The function\cite{Bal}  
$D(\beta)= \ln(\chi) - (2-\eta)\rm ln(\xi)$ and its first derivative
are also interesting  indicators of the value of $\eta$.
 Taking $\eta=1/4$, Pad\'e approximants and  DAs do not
 detect any  singular behavior of $D(\beta)$ or of its derivative
 as $\beta \rightarrow \beta_c$,
 thus confirming   
the complete  cancellation of the leading singularity in $D(\beta)$.
 Moreover, this behavior  seems to exclude the 
form eq.(\ref{momas}) of the  corrections which implies the presence of 
weak subleading singularities, while it is compatible
 with eq.(\ref{momasb}). 

In conclusion, our analysis suggests that,
in spite of their diversity, the HT extended series approach and 
the latest most extensive simulation  
 are competitive and lead to  consistent 
   numerical estimates of the highest accuracy so far possible.  

\section{Acknowledgements} We thank Prof. Ralph Kenna for a useful 
 correspondence. This work was partially supported by the italian
 Ministry of University and Research.


\end{document}